# Multi-color correlative light and electron microscopy using nanoparticle cathodoluminescence

D.R. Glenn[1], H. Zhang[1], N. Kasthuri[2,3], R. Schalek[2,3], P.K. Lo[4], A.S. Trifonov[1], H. Park[4], J.W. Lichtman[2,3], R.L. Walsworth[1,3,5]

*[1.] Harvard-Smithsonian Center for Astrophysics, Cambridge, MA 02138, [2.] Department of Molecular and Cellular Biology, [3.] Center for Brain Science, [4.] Department of Chemistry, [5.] Department of Physics, Harvard University, Cambridge, MA 02138*

## Abstract:

Correlative light and electron microscopy promises to combine molecular specificity with nanoscale imaging resolution. However, there are substantial technical challenges including reliable co-registration of optical and electron images, and rapid optical signal degradation under electron beam irradiation. Here, we introduce a new approach to solve these problems: multi-color imaging of stable optical cathodoluminescence emitted in a scanning electron microscope by nanoparticles with controllable surface chemistry. We demonstrate well-correlated cathodoluminescence and secondary electron images using three species of semiconductor nanoparticles that contain defects providing stable, spectrally-distinguishable cathodoluminescence. We also demonstrate reliable surface functionalization of the particles. The results pave the way for the use of such nanoparticles for targeted labeling of surfaces to provide nanoscale mapping of molecular composition, indicated by cathodoluminescence color, simultaneously acquired with structural electron images in a single instrument.

## Introduction:

The correlation of light microscopy with electron microscopy offers considerable scope for new discovery and applications in the physical and life sciences by providing images with both molecular specificity and nanoscale spatial resolution [1]. However, such an approach also faces substantial technical challenges in the reliable and efficient co-registration of optical and electron images [2] [3] [4]. Here, we introduce a new means of overcoming these hurdles: the simultaneous acquisition in a scanning electron microscope (SEM) of secondary electron (SE) images that are spatially well-correlated with optical cathodoluminescence (CL) from robust nanoparticles (NPs) containing stable, spectrally-distinct luminescent defects and having controllable surface chemistry. We demonstrate well-correlated NP-CL and SE images with nanoscale resolution using three species of semiconductor NPs that provide stable CL in distinguishable colors at room temperature. We also show that CL-emitting NPs can be reliably surface-functionalized, which will enable targeted labeling of molecular constituents in thin sections or on surfaces to provide multi-color nanoscale mapping of molecular composition, well-correlated with SE structural images.

The interaction of keV electrons with a solid can produce CL photons [5], a phenomenon widely used for spatially-resolved characterization of semiconductors and insulators. For imaging biological samples, the potential of CL to provide molecular localization has been recognized for some time [6]. However, efforts to obtain nanoscale CL image resolution have been hindered by low photon count rates and rapid signal degradation due to the destruction of biomolecules and organic fluorophores under electron beam irradiation [7] [8]. As described in this article, these problems may be overcome with correlated CL and SE imaging of surface-functionalizable NPs that emit stable, spectrally-distinct CL, enabling use of the CL color to identify specific molecular targets in a sample labeled by NPs. Similar methods could also be applied to the nanoscale characterization of complex, chemically active surfaces (e.g., biomaterials surfaces [9], nanocatalysts [10] or novel organic photovoltaic materials [11]). Furthermore, in conjunction with evolving techniques for wet electron microscopy [12], NP-CL could provide multi-color nanoscale particle tracking with applications in biology, colloid science, and microfluidic device characterization.

Electrons incident on semiconductors undergo a series of elastic and inelastic scattering events, depositing energy in a roughly spherical volume with characteristic radius ~1 μm for typical material densities and electron beam energies ~10 keV [12] [13] [14]. Empirically, inelastic interactions generate an average number of electron-hole pairs per incident electron $n_{EHP} \approx E_{beam}/(3E_{gap})$, where $E_{beam}$ is the beam energy and $E_{gap}$ is the semiconductor bandgap energy. Electron-hole pairs recombine radiatively by direct, excitonic or impurity-assisted processes, or non-radiatively via phonon interactions or surface recombination. In particular, recombination at color centers and other defects produces CL photons at highly characteristic wavelengths, with a spectrum and intensity controllable by doping or implantation. While much attention has been given to the CL properties of rare-earth-doped nanophosphors for application in particle detectors and display devices (e.g., see [15] [16] [17]), there have been few CL studies of well-dispersed NPs, and of nanodiamonds in particular [18].

## Results:

**Correlative imaging with spectrally distinct NPs:** We investigated NP-CL properties and collected correlated CL and SE images using a field emission SEM (JEOL JSM-7001F) outfitted with a spectrally-selective, PMT-based CL detection system (Fig. 1a). We identified three types of semiconductor NPs that provide bright, stable CL with distinct emission spectra at room temperature (Fig. 1b): (i) Nanodiamonds containing nitrogen-vacancy (NV) centers produce red CL at wavelength $\lambda \sim 620$ nm. These type 1b HPHT nanodiamonds have ~100 ppm nitrogen impurities, and were irradiated with He ions and annealed to promote NV formation (purchased from Academia Sinica,Taiwan)[19]. Dynamic light scattering (DLS) measurements gave a mean particle size of 82 ± 22 nm in an aqueous suspension. (ii) Cerium-doped Lutetium-Aluminum Garnet (LuAG:Ce) nanophosphors produce green CL at $\lambda \sim 510$ nm (Boston Applied Technologies). DLS measurements gave a mean particle size of 37 ± 13 nm. (iii) Nanodiamonds with high concentrations of 'band-A' defects generate blue CL at $\lambda \sim 420$ nm. These type 1a natural nanodiamonds (Microdiamant AG) had a DLS-measured mean particle size of 48 ± 14 nm. The mechanism of 'band-A' CL generation in diamond films is commonly associated with physical defects such as dislocations and twinning in the diamond lattice [20] [21]. We note that the CL spectra generated by all types of NPs studied here are qualitatively consistent with those of bulk samples and thin films containing the same defects.

Correlative NP-CL and SE imaging combines the best features of both multicolor optical fluorescence and high-resolution electron microscopy. To illustrate this, we acquired SE and multi-color CL images with nanoscale resolution, simultaneously and in the same instrument, for each of the three types of NPs (Fig. 2). For comparison, we also imaged each sample region using a confocal fluorescence microscope. The SE images have high spatial resolution (limited by e-beam diffraction to approximately 5 nm for our geometry), but are effectively monochromatic. The confocal images show distinct colors in fluorescence (essentially the same colors as for CL), but with photon-diffraction-limited resolution ($d_{FWHM} \sim \lambda$ / (2 NA) ~ 200-300 nm), which is insufficient to resolve individual NPs. The NP-CL images, however, provide unique information, allowing both (i) spectral discrimination between NP species, and (ii) image resolution of particles as small as ~30 nm. For many applications NP-CL will provide a considerable improvement over conventional EM techniques such as molecular labeling using gold nanoparticles, because the spectral distinguishability of the three NP markers allows spatial relationships between different molecular targets to be determined at nanoscale resolution. (Furthermore, the set of distinct NP-CL colors could be expanded by incorporating more than one type of defect in a NP, or by imaging at cryogenic temperatures to obtain narrower spectral lines [22].) Importantly, while some progress has been made in the past by simultaneously marking distinct epitopes with different sizes [23] or shapes [24] of high electron-contrast NPs, the color separation of NP-CL is unambiguous and therefore constitutes a better labeling strategy. But perhaps the greatest advantage of NP-CL imaging over previous approaches is that the CL signal can be ascertained at any resolution, whereas labeling with electron-scattering particles requires sufficient resolution to detect and/or discriminate between them. NP-CL is, in this respect, a truly multi-scale imaging technique.

**Spatial resolution of NP-CL imaging:** The limiting spatial resolution of NP-CL with scanning e-beam excitation is determined by the larger of either beam spot diameter or NP size. This is because the carrier recombination length in semiconductors like diamond is on the order of 100 nm to 1 $\mu$m at room temperature [25]; hence electron-hole pairs induced by the e-beam fill the entire NP volume and can excite CL from color centers irrespective of its point of impact on the NP. The spatial resolution of the NP-CL imaging demonstrations presented here is set by NP size (>10 nm), rather than the diffraction-limited e-beam spot size (~5 nm). To quantify the fraction of NPs of each type that produce detectable rates of CL as a function of particle size, we collected and analyzed several sets of co-localized SE and CL nanoparticle images. Characteristic data for a sample of green-CL LuAG:Ce NPs are presented in Figs. 3a-b. A high degree of correspondence between the SE and CL images is clearly visible, although some of the smallest NPs were not detectable in CL at practical integration times (~90 $\mu$s / pixel). The halo-like effect visible around the larger particles in the CL image is attributed to high energy electrons that are scattered at low angles as the e-beam scans over nearby regions of the silicon substrate, and pass into the NP volume to generate CL. Size statistics obtained for a total sample of ~1100 green-CL NPs are shown in Fig. 3c, with black bars indicating the overall distribution of NP diameters (determined from SE images) and green bars representing the sub-population with detectable levels of green CL under our imaging conditions. This CL-bright fraction comprised ~0.42 of the total NP population, with a mean bright NP diameter of 54 $\pm$ 17 nm. Similar results were obtained for red-CL (blue-CL) NPs, with a bright fraction of ~0.27 (~0.29) and mean bright NP diameter of 81 $\pm$ 25 nm (77 $\pm$ 16 nm). We draw particular attention to the five NPs indicated by red boxes in the SE and CL images of Figs. 3a-b. These NPs have diameter $\leq$ 35 nm yet also generate strong CL (> 5$\sigma$ above background). Such small, bright NPs can be selectively isolated and accumulated by centrifugal fractionation to provide NP-CL imaging resolution comparable to state-of-the-art super-resolution optical imaging with organic fluorophores [26] [27].

**NPs have good CL stability:** For bioimaging applications, the resistance of semiconductor NPs to damage under e-beam irradiation at keV energies is a key advantage compared to CL-emitting organic molecules. For each of the three types of semiconductor NPs, we demonstrated good luminescence stability by performing repeated CL imaging scans over a field of NPs deposited on a silicon wafer (Fig. 3d-e). Even after 10 full scans, corresponding to a total dose of approximately ~$10^9$ electrons for a 50 nm particle, the CL signal for each type of NP showed only minor (~ 20 - 30%) decrease. Furthermore, we found that this fractional CL signal decrease after 10 scans remained the same when the scan area was decreased by a factor of 4 while keeping the beam current and total exposure time fixed. These observations are consistent with hydrocarbon contamination of the sample surface [28] (which empirically depends on total exposure time), rather than e-beam damage to the NPs (which should increase with total electron dose per unit area). In comparison, CL emission from typical organic compounds decays by an order of magnitude after an electron dose ~100 times smaller than we applied to the semiconductor NPs [8].

**Functionalization of NPs:** Many of the applications we envision for correlated NP-CL and SE imaging will require targeted delivery of NPs to sites of interest, necessitating control over NP surface chemistry. Techniques for covalent functionalization of nanodiamonds have been demonstrated by a variety of groups [29] [30] [31], and recent work on functionalization of YAG nanocrystals [32] should be directly relevant to LuAG surfaces. As a proof of principle of NP surface functionalization while maintaining good CL properties, we used amine chemistry to bind antibodies tagged with a red fluorophore (Invitrogen Alexa-647) to blue-CL nanodiamonds. We then deposited a sample of these antibody-conjugated NPs onto a grid-etched silicon wafer, and imaged the sample in optical fluorescence followed by cathodoluminescence (Fig. 4). These images show a high degree of spatial correlation, indicating that a large fraction (~0.71) of the NPs were successfully attached to fluorophore-tagged antibodies while still exhibiting good CL emission. Subsequent optical imaging showed almost no signal in the red channel due to degradation of the organic fluorophore under exposure to the electron beam, whereas the robust blue fluorescence was unchanged.

## Discussion:

Our results demonstrate a new approach to correlative light and electron microscopy using multi-color cathodoluminescence (CL) from semiconductor nanoparticles (NPs), which can be controllably fabricated to contain spectrally distinct color centers and defects that are stable under prolonged electron-beam exposure, have excellent spectral separation, and can be surface-functionalized to enable labeling of specific molecules and structures on a wide range of samples. In the present operational regime, the NP-CL imaging resolution is set by particle size ~ 40 - 80 nm for the three types of NPs studied here. With optimized selection of small NPs having high defect concentration, we expect NP-CL imaging resolution ≤ 30 nm will soon be available. Resolution < 10 nm may eventually be realized through ongoing improvements in fabrication of small NPs with high defect concentration [33] [34]. The speed and ease of multi-color NP-CL imaging will also be enhanced by optimization of CL optical collection efficiency and parallel imaging of different CL colors, e.g., with multiple CL detection paths or use of a broadband spectrograph.

We foresee applications of correlated NP-CL and SE imaging to nanoscale functional imaging in biological samples, e.g., in serial-SEM connectomics [35], where multi-color NP-CL could allow targeted identification of molecular markers such as neurotransmitter enzymes, postsynaptic receptor types, peptides, and calcium binding proteins that differentiate classes of neurons and synapses, and be correlated with nanoscale SE structural images of thin-slice (~30 nm) neural tissue. More generally, functionalized cathodoluminescent nanoparticles will provide a powerful new tool for correlative light and electron microscopy in the physical and life sciences, enabling both molecular localization and structural imaging at nanometer resolution, simultaneously and in a single instrument.

## Methods

**DLS Measurements:** Nanoparticles (NPs) were suspended in water or ethanol at concentrations of 0.01 - 1 mg/ml. To reduce the effects of aggregation, green cathodoluminescent (CL) LuAG:Ce particles were ultrasonicated with a probe sonicator (Branson 450) at 120 W average power for 30 minutes fractionated by repeated centrifugation to remove agglomerates; this procedure was not necessary for the nanodiamond samples. Particle size distributions were then measured using a dynamic light scattering system (Delsa Nano C).

**STEM NP-CL Spectroscopy:** NP-CL emission spectra were collected using a JEOL 2011 microscope in scanning transmission electron microscope (STEM) mode, fitted with a grating monochromator-based CL detector (Gatan MonoCL3) [36]. Samples were prepared by suspending NPs in distilled water or ethanol at a concentration of 1 mg / mL, then drop-casting onto lacy carbon TEM grids (Ted Pella 01883-F). Particles were first located and imaged in panchromatic mode with a scanning e-beam at 120 keV. Because of the high electron energy and correspondingly long penetration depth, strong CL signals were best observed from relatively large NPs (diameter ≥ 200 nm). The CL spectra were collected by fixing the beam position and scanning the grating with 50 μm slit width, giving ~0.6 nm wavelength resolution. Spectra collected at different beam positions in a given NP were repeatable, as were the spectra of different NPs in the same sample. STEM-CL spectra from these larger NPs were used in selecting interference filters for the different color channels in the SEM instrument used for NP-CL imaging.

**SEM Image Collection and Processing:** NP samples for secondary electron (SE) and CL imaging were suspended in water or ethanol at densities of 0.01 - 1 g/ml, then drop-cast onto silicon wafer substrates. The substrates were first prepared by inscribing a series of 10 μm grids using reactive ion etching to facilitate image co-localization, followed by surface oxidation in air plasma. All data were obtained using a JEOL JSM-7001F scanning electron microscope (SEM), fitted with a KE Developments Centaurus CL detector. Standard imaging conditions were 5 keV electron beam energy and 1.2 nA current. The beam was scanned with a pixel size of approximately 2 nm (except in the case of the smaller LuAG:Ce particles, for which a 1.3 nm pixel size was used) and a dwell time of 2.7 μs (90 μs) for SE (CL) images. Successive CL images were taken in each color channel by turning of the beam between scans, disconnecting the photomultiplier and changing the interference filter in the optical detection path. Each CL image was associated with a simultaneously-collected SE image of the same field. SE signals were detected using a standard Everhart-Thornley detector [28]. Image processing consisted of applying a ~5 nm Gaussian blur (significantly smaller than the minimum observed bright particle diameter) to remove high frequency noise, followed by thresholding and application of a blob-finding algorithm to determine NP coordinates and size. In cases where a blob was visible in one or more CL channels that corresponded spatially to an object in the SE image (i.e., a cathodoluminescent NP), the size of that NP was taken to be the mean FWHM diameter of the SE blob. The average intensity in each CL color channel was then used to assign a color to the NP.

**Optical Imaging:** The optical fluorescence images of the NPs shown in Fig. 2 (same samples as used in the SE and CL images in Fig. 2) were collected with a Zeiss LSM-710 beam-scanning confocal microscope, using a 0.95 NA, 100× magnification air objective. (Oil objectives afford somewhat higher resolution, but create complications due to etalon effects that arise when a coverslip is placed over the silicon substrate.) Blue nanodiamonds were excited at 405 nm (excitation intensity 80 MW/cm$^2$) and their fluorescence was detected in the range 425 - 475 nm; the excitation beam was scanned with 5.2 nm pixel size and 100 μs dwell time. Green LuAG:Ce NPs were excited at 440 nm (intensity 36 MW/cm$^2$) and detected at 480 - 630 nm, with 5.2 nm pixel size and 100 μs dwell time. Red nanodiamonds were excited at 561 nm (intensity 5 MW/cm$^2$) and detected at 520 - 735 nm, with 5.2 nm pixel size and 25 μs dwell time. The fluorescence image of Alexa 647-tagged antibodies bound to blue nanodiamonds shown in Fig. 4a (same sample as used in the CL image in Fig. 4b) was obtained using a Zeiss LSM5 wide-field microscope and a 0.95 NA, 100× air objective, with excitation by a mercury vapor arc lamp (X-Cite series 120Q) and a Cy5 filter set (Chroma series 41008).

**Nanodiamond Functionalization:** The NDs (from Microdiamant) were cleaned with concentrated $H_2SO_4$ - $HNO_3$ - $POCl_4$ (1:1:1, vol/vol/vol) solution at 85°C for 3 days. After cleaning, the NDs were functionalized with carboxyl groups by refluxing in 0.1M NaOH aqueous solution at 90°C for 2 h and subsequently in 0.1M HCl aqueous solution at 90°C for 2 h. The resulting oxidized NDs were separated by centrifugation, rinsed extensively, and resuspended in deionized water. N-(3-dimethylaminopropyl)-N'-ethyl-carbodiimide hydrochloric and N-hydroxysuccinimide were dissolved together in oxidized diamond suspension, followed by addition of poly-L-lysine (PLLs) (MW 30,000, Sigma) to the suspension. Then the resulting NDs were acrylated with succinic anhydride in order to generate terminal carboxyl acid groups. These carboxyl groups on NDs were converted into a reactive N-hydroxysuccinimide (NHS) ester intermediate using 1-ethyl-3-(3-dimethylaminopropyl)-carbodiimide (EDC) and NHS. Finally, the ND samples were covered for 2 h with a solution of the antibodies (1mg/mL) in phosphate buffer at pH 7.4.

# Figures

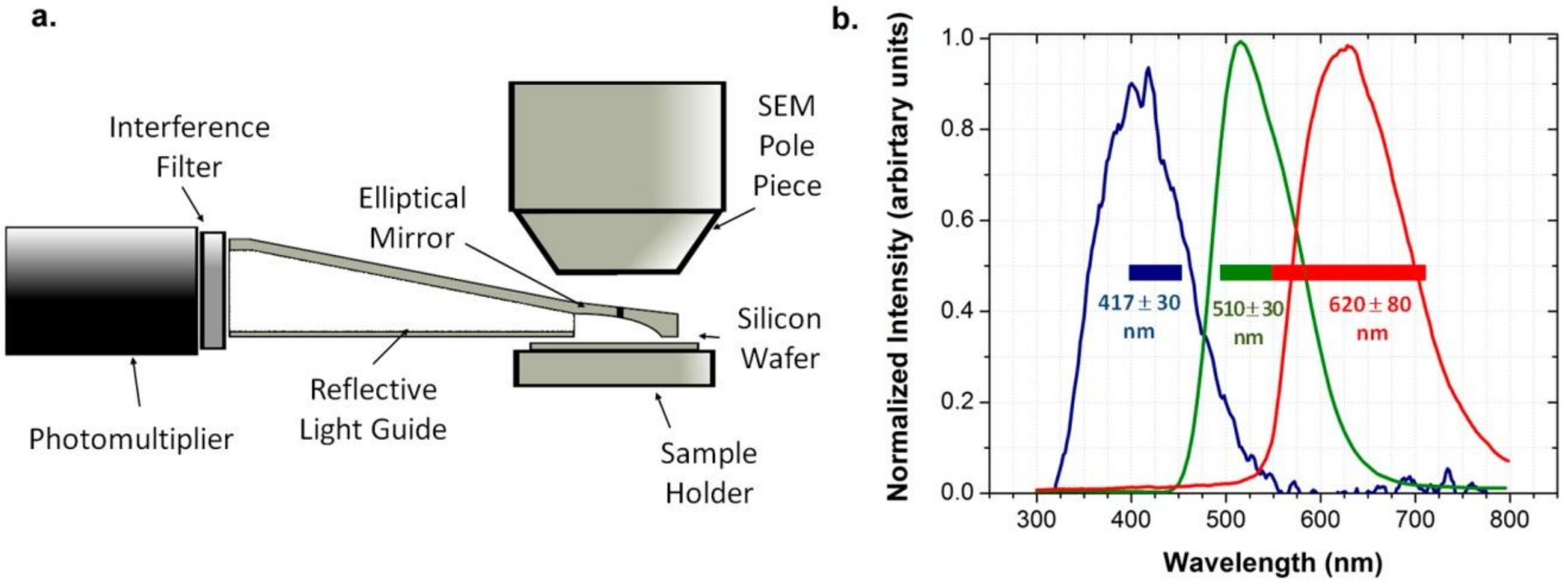

**Figure 1:** Spectrally-selective imaging of nanoparticle (NP) cathodoluminescence (CL). (a) Schematic of CL detection system integrated with a scanning electron microscope (SEM) to allow correlated NP-CL and secondary electron (SE) imaging. Samples are mounted on a silicon wafer and excited by a scanning electron beam; the resulting CL photons are collected by an elliptical mirror and directed through a light pipe onto a photomultiplier tube (PMT). Wavelength-selective optical interference filters are used to select CL light from only one NP species at a time. (b) CL spectra acquired with a scanning transmission electron microscope (STEM) for three species of semiconductor NPs: nanodiamonds implanted with nitrogen- vacancy (NV) centers produce red CL ($\lambda \sim 620$ nm); LuAG:Ce nanophosphors produce green CL ($\lambda \sim 510$ nm); and nanodiamonds with 'band-A' defects generate blue CL ($\lambda \sim 420$ nm). Normalized NP-CL spectra were used to select optical interference filters, with pass-bands indicated by colored rectangles, for multi-color CL imaging.

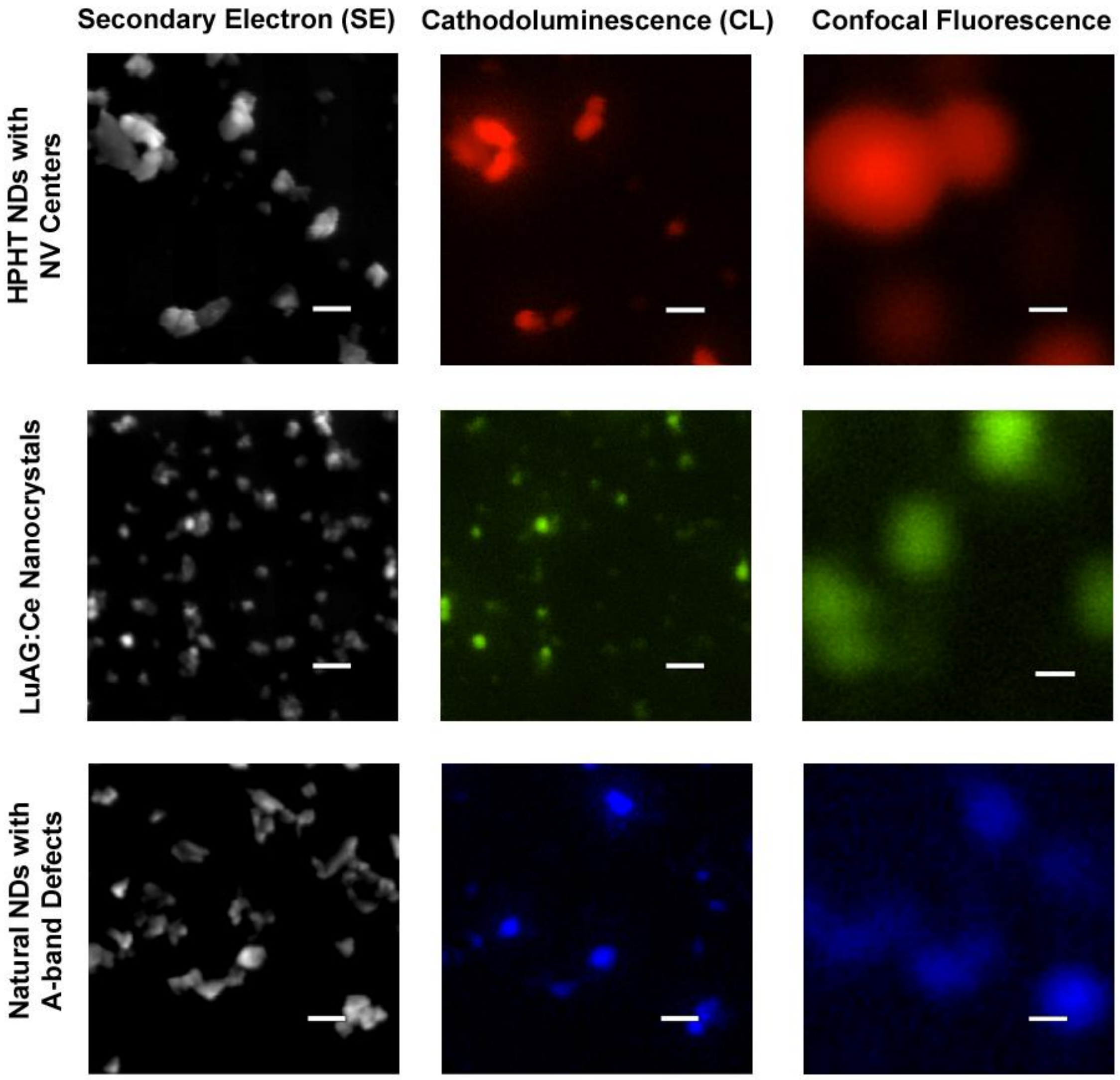

**Figure 2:** Comparison of imaging methods: secondary electron (SE), cathodoluminescence (CL), and confocal fluorescence. Each row shows images of a sample of a single NP species exhibiting (from top to bottom) red, green, and blue CL (and fluorescence) emission. Scale bars are 200 nm. SE images in the first column give excellent spatial resolution (<5 nm), but are monochromatic; whereas confocal images in the third column are in color, but diffraction-limited. CL images in the middle column are in color and provide resolution limited by NP size (~ 40 - 80 nm).

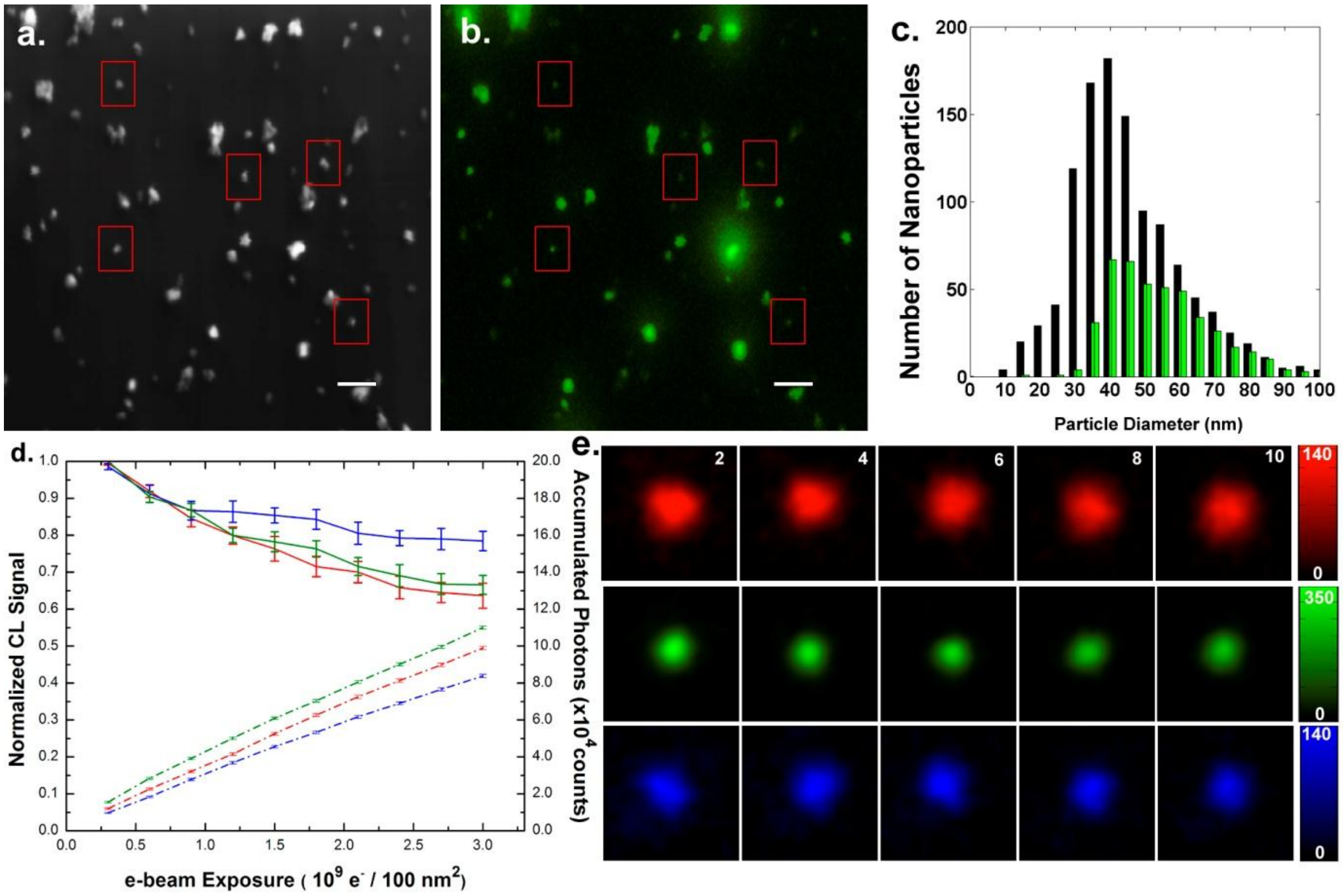

**Figure 3:** Characterization of nanoparticle (NP) size and stability of cathodoluminescence (CL) emission. (a) Secondary electron (SE) image of representative sample of green CL-emitting LuAG:Ce NPs. Scale bar is 200 nm. (b) CL image of same sample with 5 keV beam energy, 1.2 nA current, and 90 μs pixel dwell time. Scale bar is 200 nm. Red boxes indicate examples of small NPs (diameter ≤ 35 nm) exhibiting bright CL. (c) Size and CL brightness distribution for LuAG:Ce NPs. Black bars give size distribution for all NPs as measured by SE imaging; colored bars indicate NPs that produce detectable CL. Fraction of total NP population producing CL is ~0.42, with mean NP diameter of 54 nm. (d) Measured time-course of NP-CL signal, showing good CL stability for large e-beam exposure. Solid lines and associated data points (left vertical axis) give average NP-CL signal from 10 CL-emitting NPs of each species over 10 full scans of the electron beam. (Same e-beam and imaging conditions as panel (b)). NP-CL signal for each NP was normalized to its value after the first scan before averaging over NPs. Selected NPs are representative of size distribution of CL-emitting particles for each species. NP-CL signal remains ~ 70% of initial value after dose of ~ $10^9$ electrons per NP, with decrease attributable in part to accumulation on NPs of hydrocarbon contaminants [28]. In comparison, CL from organic fluorophores degrades to near-zero after dose of ~ $10^7$ electrons per $(100\ nm)^2$. Dashed lines and associated data points (right vertical axis) show total accumulated CL photon counts from a single NP of each species, selected from near the size distribution peak for that species. (Selected NP FWHM diameters: $d_{Red}$ = 97 nm; $d_{Green}$ = 51 nm; $d_{Blue}$ = 75 nm.) (e) CL images of individual NPs of each species represented by dashed lines in (d) after 2, 4, 6, 8 and 10 e-beam scans, illustrating good long-term stability of NP-CL signal intensity. Intensity scales are in units of $10^3$ CL photon counts / second / pixel.

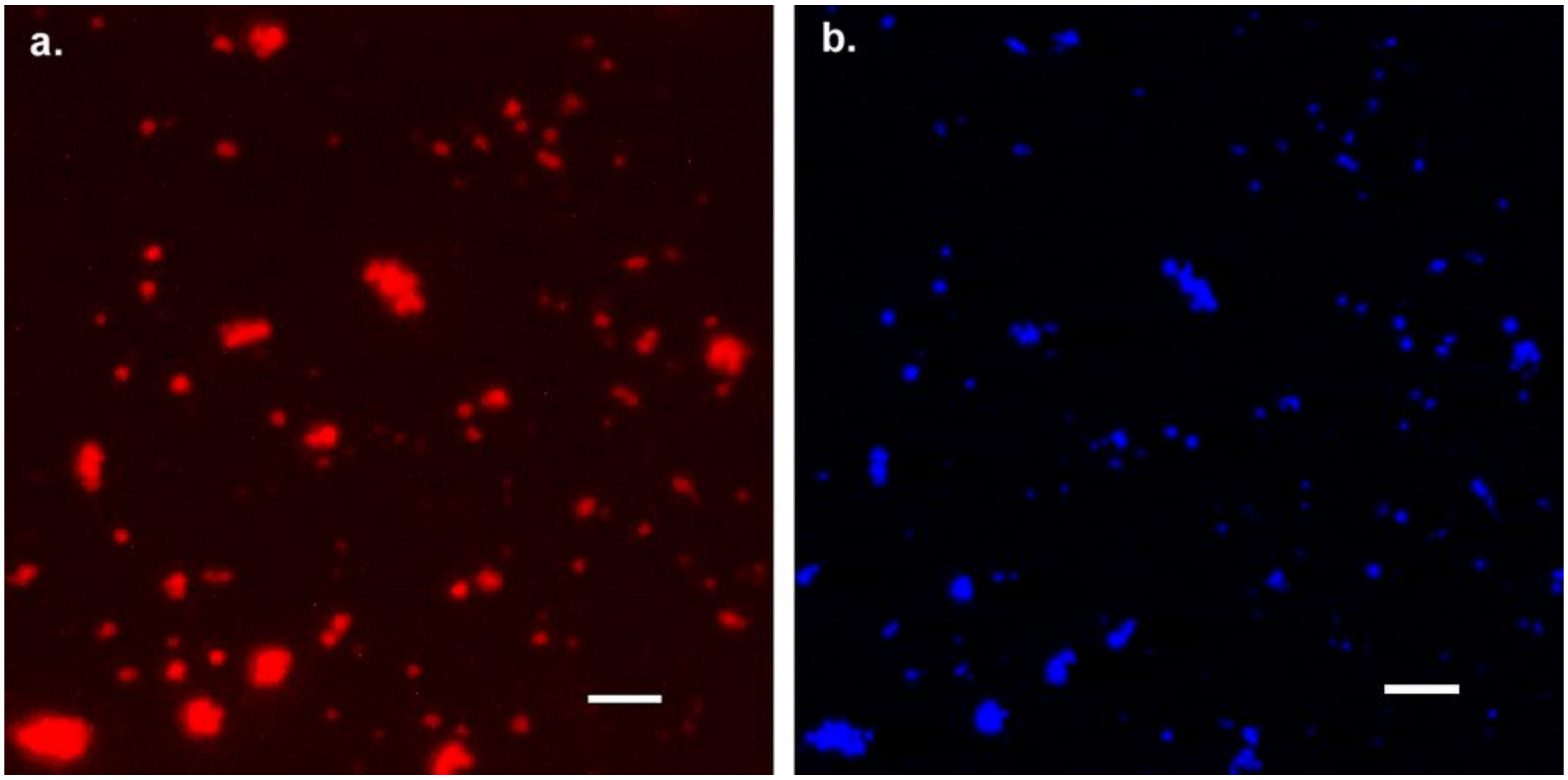

**Figure 4:** Demonstration of nanoparticle (NP) surface functionalization while maintaining good cathodoluminescence (CL) properties. Scale bars are 3 μm. (a) Red fluorescence from blue-CL nanodiamonds that were conjugated to a secondary antibody tagged with a red fluorophore (Alexa 647). Image was collected with wide-field optical microscope using red (Cy5) excitation and emission filters. (b) CL image of same field of nanodiamonds, using blue detection channel in SEM. High degree of correlation of CL and fluorescence images (~ 0.71) indicates successful NP surface functionalization.